\documentclass[a4paper,english]{spie}
\usepackage[T1]{fontenc}
\usepackage[latin1]{inputenc}
\usepackage{amsmath}
\usepackage{graphicx}
\usepackage{amssymb}
\def\puncspace{\ifmmode\,\else{\ifcat.\C{\if.\C\else\if,\C\else\if?\C\else%
\if:\C\else\if;\C\else\if-\C\else\if)\C\else\if/\C\else\if]\C\else\if'\C%
\else\space\fi\fi\fi\fi\fi\fi\fi\fi\fi\fi}%
\else\if\empty\C\else\if\space\C\else\space\fi\fi\fi}\fi}
\def\SP{\let\\=\empty\futurelet\C\puncspace }

\def\eg{{\it e.g.\/}\rm,\ }
\def\void#1{{}}

\def\lsim{~\rlap{$<$}{\lower 1.0ex\hbox{$\sim$}}}
\def\gsim{~\rlap{$>$}{\lower 1.0ex\hbox{$\sim$}}}

\makeatletter

\providecommand{\LyX}{L\kern-.1667em\lower.25em\hbox{Y}\kern-.125emX\@}
\newcommand{\noun}[1]{\textsc{#1}}

 \newcommand{\lyxrightaddress}[1]{
   \par {\raggedleft \begin{tabular}{l}\ignorespaces
   #1
   \end{tabular}
   \vspace{1.4em}
   \par}
 }

\usepackage{float}
\usepackage{graphicx}

\usepackage{babel}
\makeatother
\begin{document}

\title{New algorithms and technologies for the un-supervised reduction
of Optical/IR images}

\author{Benoît Vandame}

\maketitle

\lyxrightaddress{European Southern Observatory, Karl-Schwarzchild-Str. 2, D-85748
Garching bei Munchen, Germany.}

\begin{abstract}

This paper presents some of the main aspects of the software library
that has been developed for the reduction of optical and infrared
images, an integral part of the end-to-end survey system being built
to support public imaging surveys at ESO.  Some of the highlights of
the new library are: unbiased estimates of the background, critical
for deep IR observations; efficient and accurate astrometric
solutions, using multi-resolution techniques; automatic identification
and masking of satellite tracks; weighted co-addition of images;
creation of optical/IR mosaics, and appropriate management of
multi-chip instruments. These various elements have been integrated
into a system using XML technology for setting input parameters,
driving the various processes, producing comprehensive history logs
and storing the results, binding them to the supporting database
and to the web. The system has been extensively tested using deep
images as well as images of crowded fields (\eg globular clusters,
LMC), processing at a rate of 0.5~Mega-pixels per second using a DS20E
ALPHA computer with two processors.  The goal of this presentation is
to review some of the main features of this package.

\end{abstract}

\keywords{Image Processing, Astrometry Registration, Fringes, XML.}

\section{Introduction}

The ESO Imaging Survey (EIS) \cite{cit:luiz_mining} project at ESO is an ongoing effort to
conduct public imaging surveys and to develop the required
infrastructure and software tools to support the large volume of
optical/infrared data expected from such surveys, using dedicated
imaging telescopes such as VST and VISTA.

Since 1997 EIS \cite{cit:luiz_eis} has conducted a variety of medium-size optical/infrared
surveys. The challenge has been to cope with the variety of imagers
and observing strategies and, more recently, with the large
volume of data coming from the wide-field imager (WFI) mounted on the
2.2 m telescope at La Silla.  In the early phases of this project the
main effort was to adapt existing software to the project's particular
needs leading to the use of different packages (\eg IRAF,
ECLIPSE). This made it difficult to develop a common environment for
operations. Therefore, for the past two years an effort has been made
to develop a common, end-to-end, integrated system capable of
supporting un-supervised reduction. From the EIS survey system it is
possible to prepare observations, retrieve raw data and convert these
to science grade products in the form of fully calibrated pixel maps,
catalogs and other derived products \cite{cit:martin}. This effort has progressed in
several fronts and includes the development of an administrative
wrapper using Python, graphic user interfaces (Tcl/Tk), the
integration of a supporting database (Sybase), a suite of survey tools
and an image processing package.

Coping with the large volume of data expected in the future, and the
multi-wavelength nature of the public imaging surveys currently being
conducted requires the image reduction software to have a
high-throughput and to be instrument-independent. This has been
accomplished using a new, C-based library which integrates XML
technology and a host of new algorithms to deal with a variety of
situations encountered in observations covering the optical and
infrared domains, while using single- and multi-chip instruments.
Here a brief review of the main concepts and algorithms is presented.

\section {Preparation Steps for Reduction}

As stated earlier, the software developed is capable of supporting
single-CCD or multi-CCD instruments, and a variety of electronic
layouts (several over-scan areas and read-out ports). To support an
instrument the system requires the following information. First, a
list of basic keywords (seven in total) in the FITS header of the
original must be translated into those recognized by the system. All
others are re-computed as required to conform to the internal
convention and ensure consistency throughout the processing
phase. Second, information about the camera
used such as: the number of CCDs, their geometric layout in the array, and
the electronic layout of each CCD. The electronic layout is defined by
\noun{}the location of the \emph{overscan area}s, and the number,
location and properties (read-out noise and gain) of different
read-out ports. These define different quadrants on the image, which
must be known for a proper reduction. Third, the definition of
{\it Reduction Modes} (RM; see section \ref{sec:reducmode}) specific to that
instrument.

Once this information is available and the raw data are on disk the
process can start. The first step is to check the number of fits
extensions in the raw images. If this number corresponds to the number
of CCD chips expected for the instrument, the original raw image is
split into individual ones. The second step is to translate the
keywords in the fits header of the raw images, using the dictionary
mentioned above, and to check for their existence and integrity.  The
third step is to organize the raw images by recognizing their types
(science, standards, calibration) and to create reduction blocks
(RBs). These consist of groups of images taken sequentially and
sharing the same filter and/or position (in the case of science
images).  Calibration frames are also sorted by their type namely
bias, dark, dome flats and twilight flats. Dark exposures with the
same integration time are grouped together. These calibration RBs can
be further combined according to a period of validity defined by the
user.  Finally, each RB is assigned a RM which tells the system how
these images should be reduced and the parameters to use for each
sub-process.

The reduction mode carries information about the instrument (number of
chips, chip size, overscan region, read-out ports, gain,
readout-noise) and attributes such as the type of image (dark, bias,
flat, standard, science), type of survey (extragalactic or galactic
fields), and passband (optical/infrared). Each of these attributes may
require a special sequence of processes and/or suitable parameters.
The information contained in the RB is used to define the sequence of
processes and the optimal parameters for each instrument and
passband. The present system is capable of sorting out these various
pieces of information and use the specific modules required with the
appropriate set of parameters. To a large extent this is the major
advantage of an integrated system as the one described here.

For science frames, besides the RB, one has also to provide additional
information such as the astrometric reference catalog to be used in
the registration of the images and the definition of a reference grid
onto which the final reduced images will be mapped. This reference
grid is defined by a reference sky position, a pixel scale and
orientation (\eg north-up; east-left) and a projection (\eg TAN, COE).
The use of a reference grid is not mandatory but is recommended for deep
observations and mosaics since it is easier to co-add images sharing
the same grid convention.

In the survey system version, the Python wrapper can use the original
groups and apply additional constraints to filter out images. For
instance, it is possible to restrict images based on the variation of
the seeing (currently based on the values of the seeing monitor), the
time or airmass interval covered, and enforce that the reduction block
corresponds to the original observation blocks (OB), as defined by
ESO's data flow system.

Before the actual reduction of the data, some additional checks are
required.  The first check is on the accuracy of the reference pixel
stored in the header. This is important in order to detect possible
large offsets from the expected location which may occur due to
changes in the pointing model at the telescope. This is also critical
given the convention adopted by ESO for the CRVALs and CRPIXs values
in the FITS header of the raw images (especially SUSI2and WFI). This
check is carried out using a cross-correlation technique between two
images. The first is a low-resolution (under-sampling by a factor of 8
the original pixel scale of the CCD considered) version of a mock
image created from the reference catalog drawn from a much larger area
than one CCD (ranging from 4 to 36 times the original field-of-view).
The other is constructed from the science image to the same resolution
as the one adopted earlier using wavelet transform, padded with zeros
to match the mock image size. In preparing the low-resolution science
image, care must be taken to remove cosmic rays and the effect that a
few bright stars may have on the results. A more detailed description
of the method will be presented in a forthcoming paper.  The peak of
the cross-correlation yields the offset that must be applied to the
values of CRPIXs in the header to correspond to the values of the
CRVALs. This correction is then applied to all frames (individual
chips) belonging to the same night or an arbitrary period of validity
specified by the user.  This procedure is possible thanks to the speed
of the algorithm which works with low-resolution images and benefits
from the Fast Fourier Transform (FFT). One advantage of carrying out
this reference pixel check at the preparation phase is to assure that
during reduction only a refinement of the astrometric solution is
required. Note, however, that the method is only valid for correcting
translations. When dealing with multi-chip instruments some algorithms
(see section \ref{sec:calib}) require information about the exact
location of each chip and this must be checked by the system. Since it
is not advisable to rely in the order in which they appear in the FITS
file, which may (have) change in time, this must be computed.  Once
new CRPIX values are available these are used to define the geometry
of the array and to cross-check if the computed CRPIXs are consistent.

\section{Handling Calibration Frames}
\label{sec:calib}

Normally, the first RBs to be reduced are those associated with the
calibration frames.  However, it is also possible to assign external
master calibrations, in which case this step can be skipped.  Each
calibration RB of the ``closed'' type (dark and bias) containing at
least three frames a median value is computed and a master bias and
dark are created following the validity period specified. Several
master darks may exist with different exposure times.  In general,
darks are only used for IR reduction, since their contribution for
modern optical CCDs is negligible. The RB for flats (dome or twilight)
contains only frames within a specified range of counts, the others
being rejected in the preparation phase. Each frame is then subtracted
by the master bias or, in the case of IR, by the corresponding (same
exposure time) master darks. Each frame is then divided by their mean
to provide normalized images and a median value is computed to produce
the master flat. The same procedure is adopted for dome and twilight
flats, but currently for test reductions the twilight flat is
normally used leading to good results. One exception is the IR images
taken with SOFI at La Silla. SOFI uses the so called special-flats,
requiring a special reduction procedure which is currently not yet
available. Instead the flats prepared on the mountain are used.

Once a master flat is available a bad-pixel (hot/cold/dead pixels) map is
created by flagging pixels with intensities greater or smaller than a
threshold level on an image storing the difference between the master
flat and a smooth version of it produced by computing a median in a
10$\times$10 box. This is done automatically and on-the-fly for each
reduction with no need for saving this map. At this point the
information about the read-out ports is important in order to avoid smoothing
over the border of different quadrants. Otherwise these borders will
be flagged as bad-pixels.

The master flat also characterizes the noise level of the pixels which
can be used to define the notion of weight map (pixels with a low
efficiency should have a smaller signal-to-noise). This notion is
critical in the co-addition of images and for a homogeneous source
extraction at a fixed signal-to-noise level. 
The weight map is equal to the master flat. It can also be
generalized by multiplying it by the bad-pixel map which gives zero
weight to the rejected pixels. This weight map is then associated to
all raw science frames within the validity period of the master flat.
It is important to emphasize that the weight map is equivalent to an
effective exposure map.

The procedure described above also applies to multi-CCD instruments,
with the system splitting the raw images into individual CCDs. The
administration of these several files is carried out internally by the
system and it is transparent to the user. However, for CCD arrays an
additional step is necessary, namely the correction for variations of the
gain from chip-to-chip along the array so as to enforce that all the
master flats of the different CCDs have the same gain.

This correction is computed separately and for each filter using long
exposures and assuming that all CCDs in the array have the same
gain. Hence, after they are corrected for instrumental effects (bias,
dark, flat), they should share the same background value.  Under this
assumption the background is computed at the four borders (left, top,
right, bottom) of each CCD in the array. If differences in the counts
are found, a flux-scaling factor for each CCD is computed such as to
minimize these differences. The flats are then multiplied by the
derived flux scale. For the instruments considered (\eg SUSI2, WFI)
these differences are typically less than 10\%.  It is important to note that this
calculation does not take into account nor corrects for illumination
effects. The derivation of the illumination correction needs to be
done separately and it is not yet fully implemented.

In principle, this procedure prevents the need for the calculation of
independent photometric calibrations for each chip. However, the
robustness of the procedure  must be independently monitored by
computing photometric solutions chip by chip. Such a facility is
provided by the survey system.

\section{Handling Science Frames}
\label {sec:reducmode}
\subsection{Reduction}

The pipeline offers different ways to process a reduction block of raw
images.  As mentioned earlier these are set by the \emph{Reduction
Mode} (RM) assigned to it. In the present framework four main modules
are available as schematically shown in
Figure~\ref{fig-reduc-mode-general}. Also indicated at the bottom are
the external dependencies (\eg master calibration images and chip
layout, weight map, and reference catalogs for the absolute
astrometric solution) of each of the modules. The current system
consists of the following steps:

\begin{figure}
\begin{center}\includegraphics[  height=4cm]{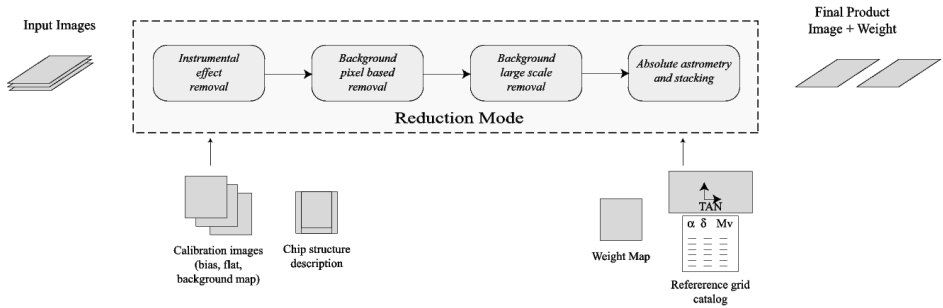}\end{center}

\caption{\label{fig-reduc-mode-general}Main available steps of a Reduction
Mode.}
\end{figure}

\begin{itemize}

\item \noun{Removal of instrument signature} - For each optical image
the master bias is overscan corrected quadrant by quadrant and
subtracted from the science images. Next the science images are
divided by the master flat. A similar procedure is adopted for the IR
images except that instead of the master bias, the master dark with
the appropriate exposure time is used. An option is available to
interpolate across bad pixels. In general, however, since multiple
exposures are available this is taken into account by the weight map
during the co-addition process.

\item \noun{Removal of local background} - this module is normally
used to deal with images observed in passbands where both low and
high-frequency features such as fringes are present in the
background. This is typically the case for the red passbands in the
optical (R and more conspicuously in I) and in the infrared domains.
A first crude attempt to estimate the local background is done by
assuming that the observed features are stable in pixel coordinates.
A first estimate of the local background level is computed as the
median of the the intensity of all images in the RB at that
pixel. Next all images in the RB are subtracted by this estimated
background. However, as shown in sub-section \ref{sub:Fringing-removal}
this estimate is bias and greatly affects the flux of faint
objects. To improve the background estimate it is necessary to mask
out objects. In order to do that the images are registered using one
of them as reference, and the values are added and divided by the number of
images contributing at the position of the pixel in the new
grid. Objects are extracted from this image and a mask image is
created. This mask is then applied to the original images in the RB at
their original pixel coordinates. Pixels inside the masks are then
discarded in the calculation of the mean for a second estimate of the
background. This second estimate is then used to subtract the
background of each image. The resulting RB is then used by the next
module.  In cases where there are not enough exposures or for very
crowded fields the system uses an external fringing map as discussed
in sub-section \ref{sub:Fringing-removal}.

\item \noun{Removal of low-frequency background} - this module
computes and removes the background of each image. This procedure
assumes that the images have a smooth background (using a median of a
top-hat filter) and that the smoothing scale is larger than the
largest object in the image.  For each image a mini-background is
computed (according to the smoothing scale adopted) and an estimate of
the background, for each quadrant, is computed merging a user
specified number of mini-backgrounds. Once this estimate is available
the mini-backgrounds are re-sampled back to the original image scale
and subtracted from all images in the RB. In the case of IR
observations carried out in nights of variable conditions, it is
recommended to deal with each image separately. At this point any
residual imprint of the quadrant structure in the CCD is removed.

\item \noun{RB stacking \& Astrometric calibration} - the final module
of the reduction part is responsible for producing a final image which
combines all the images in the RB, normalized to one second
integration time (and computing an effective gain), by calculating the
mean value at each location, and astrometrically calibrating the final
image. The noise of the normalized images in the RB are computed and
for RBs with at least three images the noise distribution is examined
and outliers are removed. In case the individual images have
sufficient high S/N objects, the astrometric reference catalog
specified in the previous section is used to create a mock image (as
discussed above) using the reference grid specified at the preparation
level, to determine the absolute astrometric registration for each
image which are warped and then combined. This is normally the case
used for optical images with exception perhaps of usually shallower
$U$-band exposures. For the latter and for all IR images a relative
astrometric solution is found, images are then warped and
combined. This final image is then astrometrically calibrated using
the reference catalog mapped (warped) to the final reference grid,
exactly in the same way as above. Once the images in the RB are
registered (absolutely or relatively) cosmic rays are removed using a
sigma-clipping technique.  At this point, the Hough transform (see
sub-section \ref{sub:hough}) is applied to identify possible satellite tracks. If found,
pixels along a strip surrounding the track are rejected. Finally, the
weight map is used to mask out bad pixels.  This is done before
warping the images. In general, this module deals with sky-subtracted images coming
out from the previous modules. However, this is not required since in
some cases it is of interest to keep the smooth part of the original
image background. On the other hand, this module can only be used
after fringes have been removed, since they greatly impact the wavelet
decomposition, specially in the medium-resolution scales which is used
for the first estimate of the astrometric calibration. In the case of
images for which the background has not been removed, offsets in the
background level, due to variations in sky brightness during the time
interval of the RB, are corrected for using one of the images as
reference.  The final stacked image is trimmed according to a
user specified minimum number of contributing images. The weight map
of the final image is the sum of the original weights properly
normalized and warped, taking into account the exposure times of the
original images. 

\end{itemize}

Typical examples of reduction modes are: 1) deep IR images and optical
images in I-band all modules are active; 2) in the case of the bluer
passbands, the local background module can be bypassed; 3) in the case
of crowded fields, such as those in the Pre-Flames survey, only the
first and the last modules are used.

\subsection{Final Products}

Often the observations are carried out as a sequence of
OBs. Therefore, at the end of the reductions several reduced RBs are
available all sharing the same reference grid. At this point these
intermediate images can be combined using the weight maps and the
noise, as weight. At this stage a further rejection of cosmic rays is
applied leading to the removal of lower S/N features not detected in
the original frames. In the case of multi-chip instruments at the end
of the process the RBs of the individuals CCDs are merged together
into the final image as specified by the reference grid. This assumes
that flux scaling correction of the flats described earlier is
adequate.  Otherwise, an external correction factor for each CCD has
to be provided and an external module has to be used to combine and
merge the different chips.

\section{Algorithms}

\subsection{Removal of Fringes}
\label{sub:Fringing-removal}

As mentioned in the previous section fringes are a common phenomenom
present on near-infrared images. Fringes are an additive signal
stemming from sky emission lines, making a non-negligible contribution
to the total signal, typically of the order of 10\% of the background
level. Hence, subtracting its contribution is essential for a proper
reduction of the image. While in the optical, the fringing pattern is
relatively stable (but exceptions exist) over the time scale of a
night, in the infrared the fringes vary very quickly. Fortunately, the
fringing pattern is largely independent of the pointing and
consecutive images share the same pattern. This allows them to be
separated from the astronomical objects by shifting the pointing
observations. While the objects will fall in different pixel locations
the fringes will not.

\begin{figure}
\begin{center}\includegraphics[  height=4.6cm]{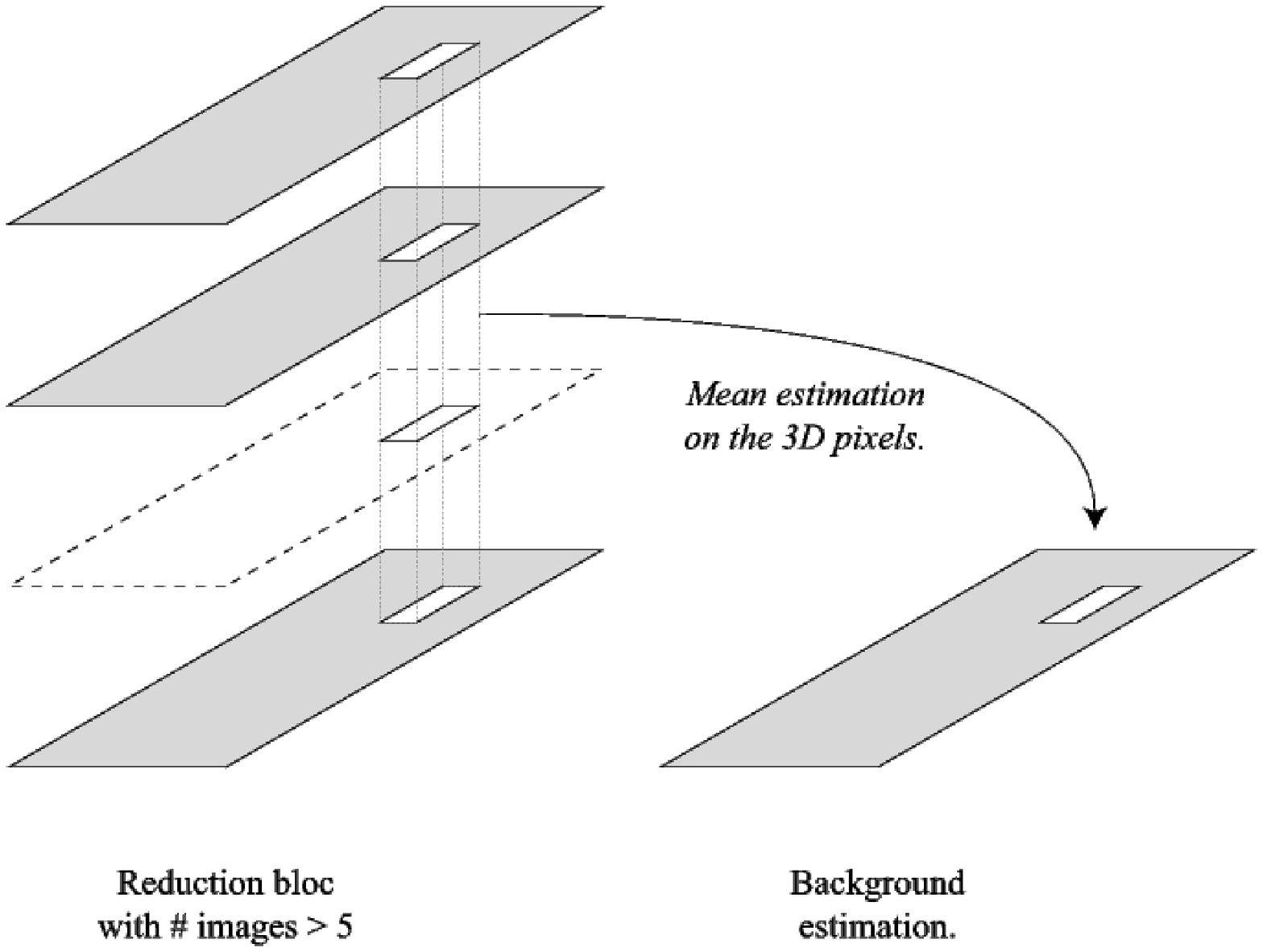}~~~~~~~~\includegraphics[  height=5cm]{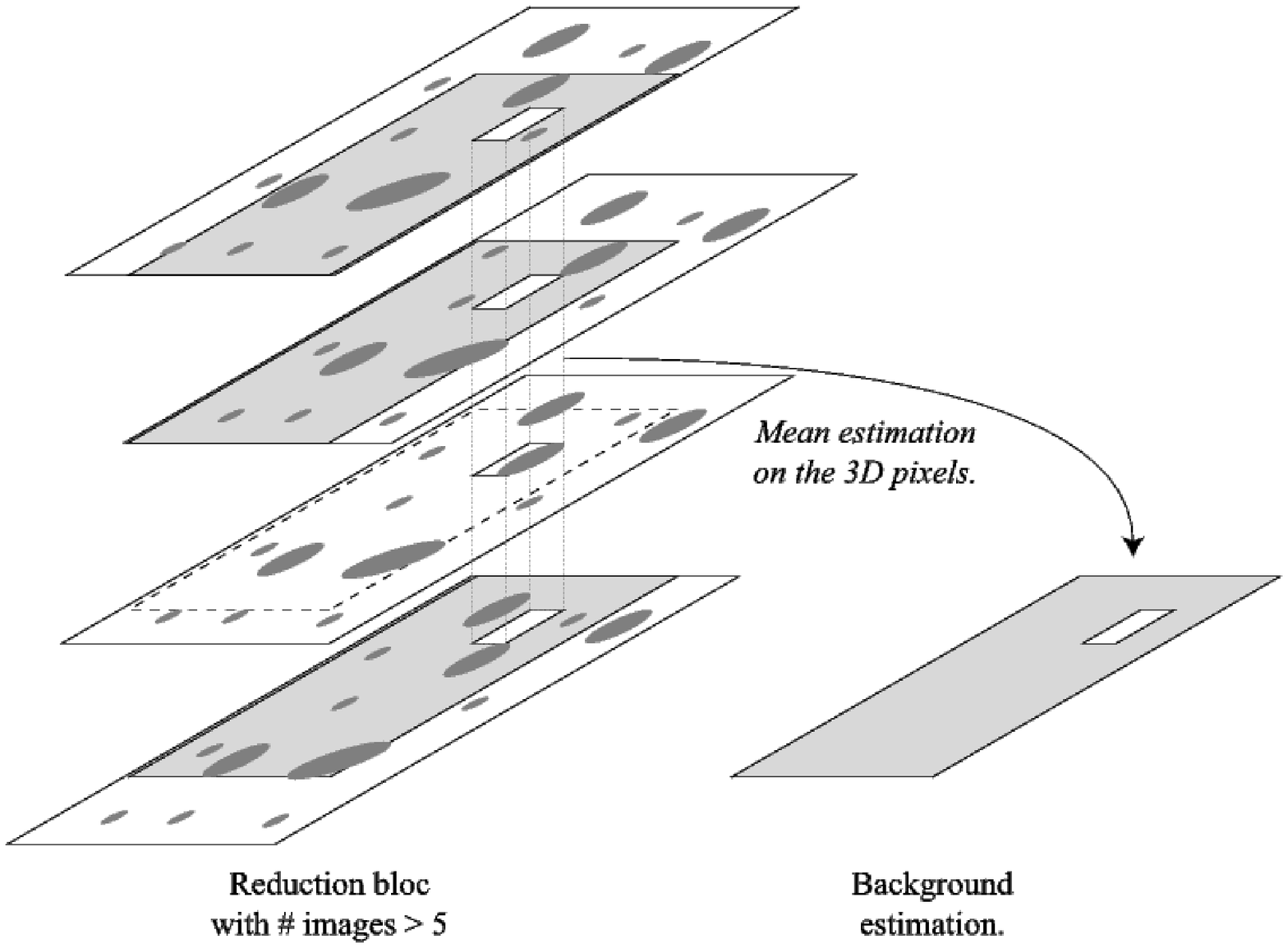}\end{center}

\caption{An illustration of the two methods used  to estimate the
background: Left: \label{fig:fringing-1} a simple median filtering;
Right: \label{fig:fringing-2} median filtering after masking objects
in the final stacked image, as described in the text.}
\end{figure}

Therefore, a first crude estimate of the fringing pattern is to
compute the median at every pixel using the  estimator $E$ given by

\begin{equation}
B_{i,\, j}\, =\, E\, \left(P_{n_{i,\, j}-}\mu _{n},\, \forall n\right)\label{equ-fringing}
\end{equation}

\noindent where $B_{i,\, j}$ is the computed background at the
location $(i,\, j)$, $P_{n_{i,\, j}}$ is the counts of the pixel
$(i,j)$ of the $n^{th}$ image, and, $\mu _{n}$ is the mean of the
$n^{th}$ image. Subtraction of $\mu _{n}$ is important because it may
vary from one image to another in the case of non-photometric nights.
The estimator apply on the RB is shown in Figure \ref{fig:fringing-1}
on the left panel.

Unfortunately, this estimator is biased due to the presence of faint
astronomical objects which cannot be detected in the individual images
due to their low S/N.  However, if all images are combined these
objects can have a significant contribution. For instance, for an RB
with 60 images objects detected in the final image with a S/N $\sim$ 2
have a S/N $\sim$ 0.25 on the individual images. Neglecting their
contribution in estimating the background leads to an overestimate of
the background, and consequently to an underestimate of the flux of
the objects. This effect can be as large as 0.2~mag near the limiting
magnitude.

In order to improve the estimator $E$ in equation~(\ref{equ-fringing})
it is necessary to distinguish pixels which have the
{}``\emph{background value}'' from those contaminated by the light of
objects as illustrated in the right side of Figure
\ref{fig:fringing-2}. In order to do that one has to use the first estimate
of the background, combine the sky-subtracted images, detect objects,
and create masks which encompass surrounding pixels containing the
object. These masks are then mapped back to the individual images and
used to discard pixels in a second estimate of the
background. Typically, the mask size is taken to be about twice the
estimated size of the objects.
The estimator apply on the RB with the mask is shown in Figure \ref{fig:fringing-1} on the right pannel.

If the background is evolving during the total exposure time, it can
be estimated only from some images observed consecutively. The
algorithm allows to specify a range of images to be used to estimated
the background.  In the IR the background is evolving faster, and a range
of typically 15 images is used to compute the background,
corresponding to 15 minutes of observation.

One limitation of the method is that, for large objects and very dense
regions it may be impossible to ``see'' the sky, in which case holes
will appear in the final image. In order to avoid this case a proper
choice of the amplitude of the jitter pattern is critical.

\subsection{Astrometric registration using wavelet transform}

In the pipeline astrometric solutions are computed either in a
relative or in an absolute way. For instance, relative astrometry is
done for the stacking of an RB using one of the images as reference,
while an absolute astrometric calibration requires an astrometric
reference catalogue and a grid. In both cases the solution is found
using the concept of multi-resolution (reference), by comparing
low-resolution images obtained from a wavelet decomposition. In the
case of a reference catalog this is done by first creating a mock
image to map the objects listed in the catalog onto a reference
image according to the grid convention. This method is referred to as
pixel-based image registration \cite{cit:djamdji}.

The calculation of the geometrical distortion between the two images
requires: 1) the identification of a set of well defined astronomical objects
(control points, hereafter CP) in both images; 2) the definition of a
functional form for the deformation model which maps a point $(x,y)$
in the input image to a point $(X,Y)$ on the reference image; 3) the
creation of the distorted or corrected image, which requires the
adoption of an interpolation or warping method.

The distortion model adopted by the pipeline is a pair of bivariate
polynomials given by:

\begin{equation}
\left\{ \begin{array}{ccc}
 X_{i} & ={\displaystyle \sum _{p=0}^{N}\sum _{q=0}^{N-p}}\, a_{pq}\, x_{i}^{p}y_{j}^{q}= & P\, (x_{i},y_{i})\\
 Y_{i} & ={\displaystyle \sum _{p=0}^{N}\sum _{q=0}^{N-p}\, b_{pq}\, x_{i}^{p}y_{i}^{q}}= & Q\, (x_{i},y_{i})\end{array}
\right.\label{eq:distort}\end{equation}

Where $\textrm{(}X_{\textrm{i}},Y_{i})$ are the coordinates of the
$i^{th}$ CP in the reference image and $(x_{i},y_{i})$ the
corresponding CP from the input image. $N$ is the degree of the
polynomial. Usually, the degree of the polynomial is one or two for
instruments with a small field-of-view and up to four otherwise.
Then follows the computation of the unknown parameters ($\frac{(N+1)(N+2)}{2}$ for
each polynomial) using the least mean square estimator (LMSE).

The most difficult step of the process is to associate the CP from the
input image with the CP from the reference image. CP are randomly
spread and can be very dense. The associations cannot be done using
any \emph{window search} because multiple associations will occure and
will bias the distortion estimation. The idea of the wavelet
transform, is to work with different resolutions and therefore give a
constraint for the minimum distance between CP.

Let $I_{r}$ be the reference image and $I_{i}$ be the input image.
On each image the wavelet transform with the so called \emph{algorithm
à trous} \cite{cit:mallat} is performed up to the scale $L$ producing two wavelet
images $W_{il}$ and $W_{rl}$ with: $l\, \in \, (1,L)$ from the
two images $I_{i}$ and $I_{r}$. $L$ being the initial dyadic step,
it corresponds to the large details of the images. A 
detection procedure on $W_{iL}$ is achieved and only the local maxima with amplitude
greater than three times the sigma of the noise are kept. Theses local maxima
play the role of CP. They correspond to significant image patterns,
and have to be found in the corresponding wavelet plan of the reference
image $W_{rL}$. The association is done with a window search. The
search radius is set to $2^{L-1}$ pixels, and corresponds to the half
of the maximum wavelength (Shannon constraint) appearing at the scale
$L$. This radius assures no multiple associations. At this scale
identical CP can be matched with confidence, and therefore the
relationship between the coordinates of the different images can be determined.
This provides a rough estimation of the distortion model. Considering now
the wavelet images of order $L-1$, a new set of maxima
in each wavelet plan is detected. The coordinates of each maximum
detected are then transformed in the reference image using the previous estimation of the
distortion model. This follows an iterative process until the last scale
corresponds to the wavelet plan of the best resolution, establishing thus
the best goemetrical correspondance. A first degree polynomial is used for the scale
$L$, but the degree can increased for the last scales.

For each input pixel at location $(i,j)$ is computed $(k,l)$, where
$k=P\, (i,j)$ and $l=Q\, (i,j)$. The coordinates of the distorted
pixel $(k,l)$ are generally not integers, so an interpolation must be
done to calculate the intensity value for the output
pixel. Nearest-neighbor, bi-linear, bi-cubic, splines, and Lancoz
interpolations are the most widely used \cite{cit:swarp}. Currently,
only the nearest-neighbor and Lancoz are used with the latter
providing the best results.

\begin{figure}
\begin{center}\includegraphics[  height=50mm]{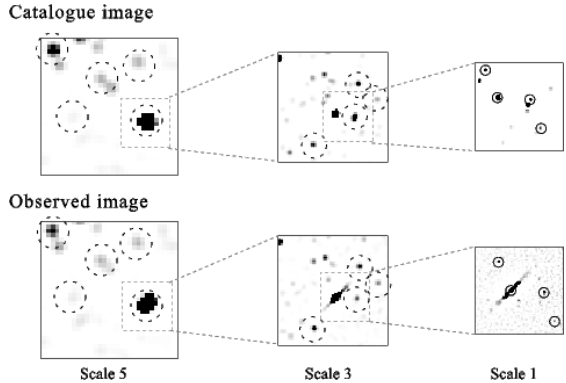}\end{center}

\caption{\label{fig:astro-example}Wavelet transform of the reference
image (simulated image from a catalog), and the wavelet transform of
the input image to be registered. The circles show some associations
done with the window search scale by scale. Only a part of the wavelet
planes are shown via a zoom from one scale to a lower one, in order to
improve the visibility. Scale 5 is the last scale (lowest-resolution)
of the wavelet transform.  A rough distortion model is deduced from
this scale. It is then refined with the other scale up to scale 1. The
final distortion model is made with a second order polynomial.}
\end{figure}

The advantage of the multi-resolution approach is that it is
insensitive to the presence of spurious objects in the reference
catalog (stars with large proper motion, objects that do not appear on
the observed image due to their color). This multi-resolution approach
has also proven to be very efficient in very crowded areas such as the
Small Magellanic Cloud. Exhaustive tests have also been performed on
images of the open cluster Messier 67 observed with wide-field imager
\cite{cit:yazan}.

\subsection{Satellite track detection}
\label{sub:hough}

A frequent problem encountered in imaging surveys using wide-field
imagers is the presence of a significant number of satellite tracks.
This is a non-trivial problem for survey work since these features,
if not identified in the original exposures, will appear in the
final co-added image.  The presence of these tracks are not only
unpleasant from the cosmetic point of view but also have a
considerable impact on the derived products (source extraction). Once
propagated to the final image, none of the available options are
adequate. One can either mask the final image, which may continue to
affect the source extraction, or to go back to the individual
exposures, mask the visible tracks, and stack the images again. While
this manual process can be done for a few images it is certainly not
feasible for large data flows given the high incidence of these
features. Therefore, a robust system must be able to identify and
remove these unwanted features automatically during the processing
phase which precedes the stacking \cite{cit:vandame_hough}.

Fortunately, this problem can be tackled using the Fast Hough
Transform (FHT) \cite{cit:hough} since a track
consists of a set of points aligned in an otherwise quasi-random
distribution of points. The method is based on the fact that the FHT
transforms points to the original space into lines (using Bresenham's
procedure) in the Hough space. In this space the intersection of the
lines correspond to the parameters of a straight line in the original
space.

\begin{figure}
\begin{center}\includegraphics[  height=4.5cm]{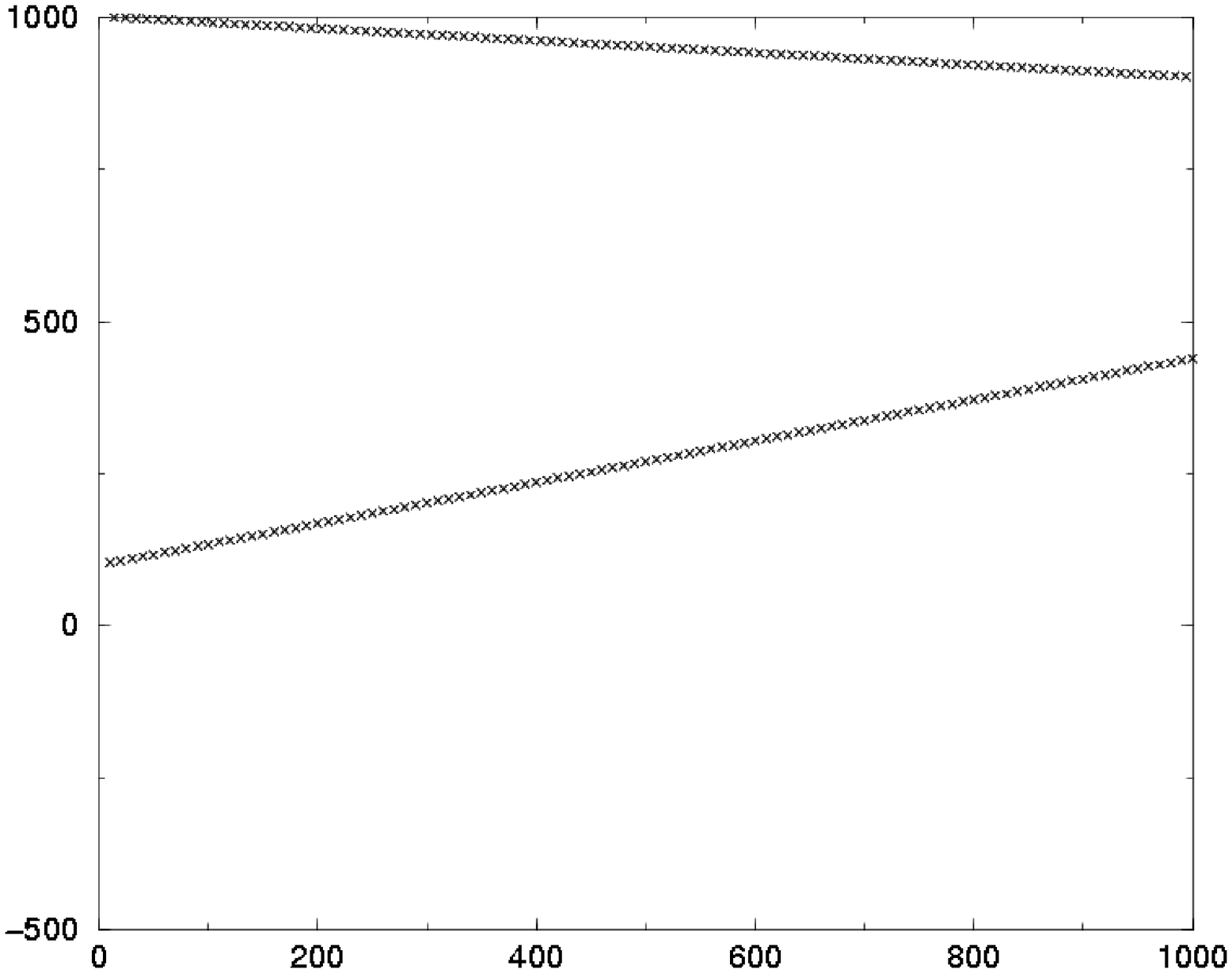}~~~~~\includegraphics[  height=4.5cm]{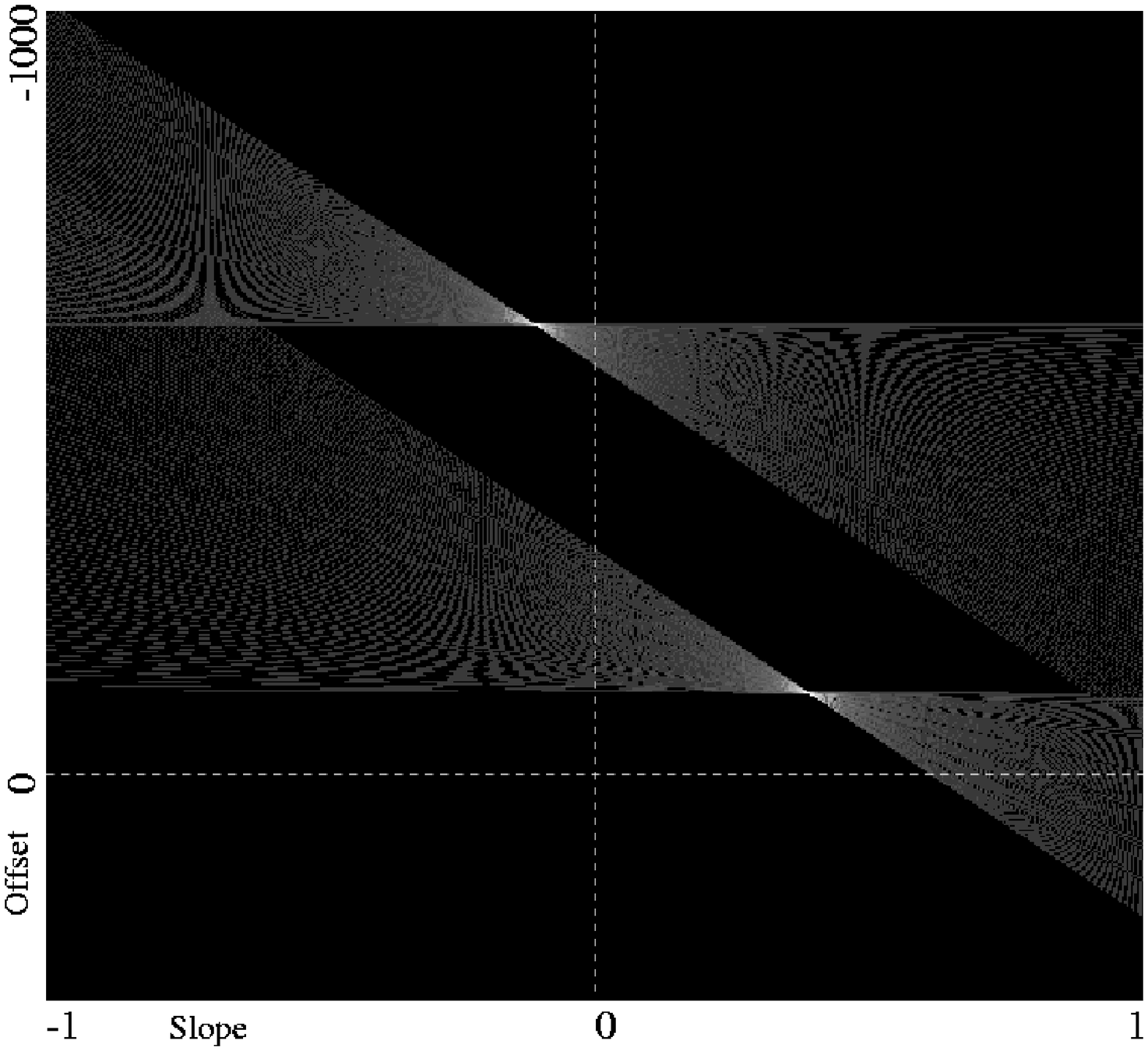}\end{center}

\caption{\label{fig:hough-ex}Example of the FHT of two sets of aligned points.}
\end{figure}

The Hough space is a two-dimensional space in which each point $x,y$
in the image space is transformed into a line given by $y=-x*X+y$. The
amplitude in the Hough space is the sum of the number of points
contributing at each position in this space. The alignement of $N$ points in real
space following the relatiion $y_{i}=a*x_{i}+b$, where $i=1,N$, will imply
a peak in the intensity of the amplitude in Hough
space at the point ($X=a,Y=b$) given by $b=-x_{i}*a+y_{i}$. This
transformation is illustrated in Figure \ref{fig:hough-ex} where two
tracks in real space are shown in the left panel and the Hough
transform of all points along these two lines are represented in the
Hough space in the right panel. As discussed above the intersection of
the point-transformed leads to an intersection in the Hough space
yielding a local maximum. The identification of these high-amplitude
peaks yields the coefficients $a$ and $b$ that correspond to the slope
and the offset of the track in real space.

Note, however, that the case shown is valid only for
slopes in the interval $[-1,+1]$. In order to deal with slopes outside
this range and limit the size of the FHT space, the original catalog
is flipped to satisfy the same conditions. Therefore, the actual implementation
of the algorithm consists of two steps.

\begin{figure}
\begin{center}\includegraphics[  height=4.5cm]{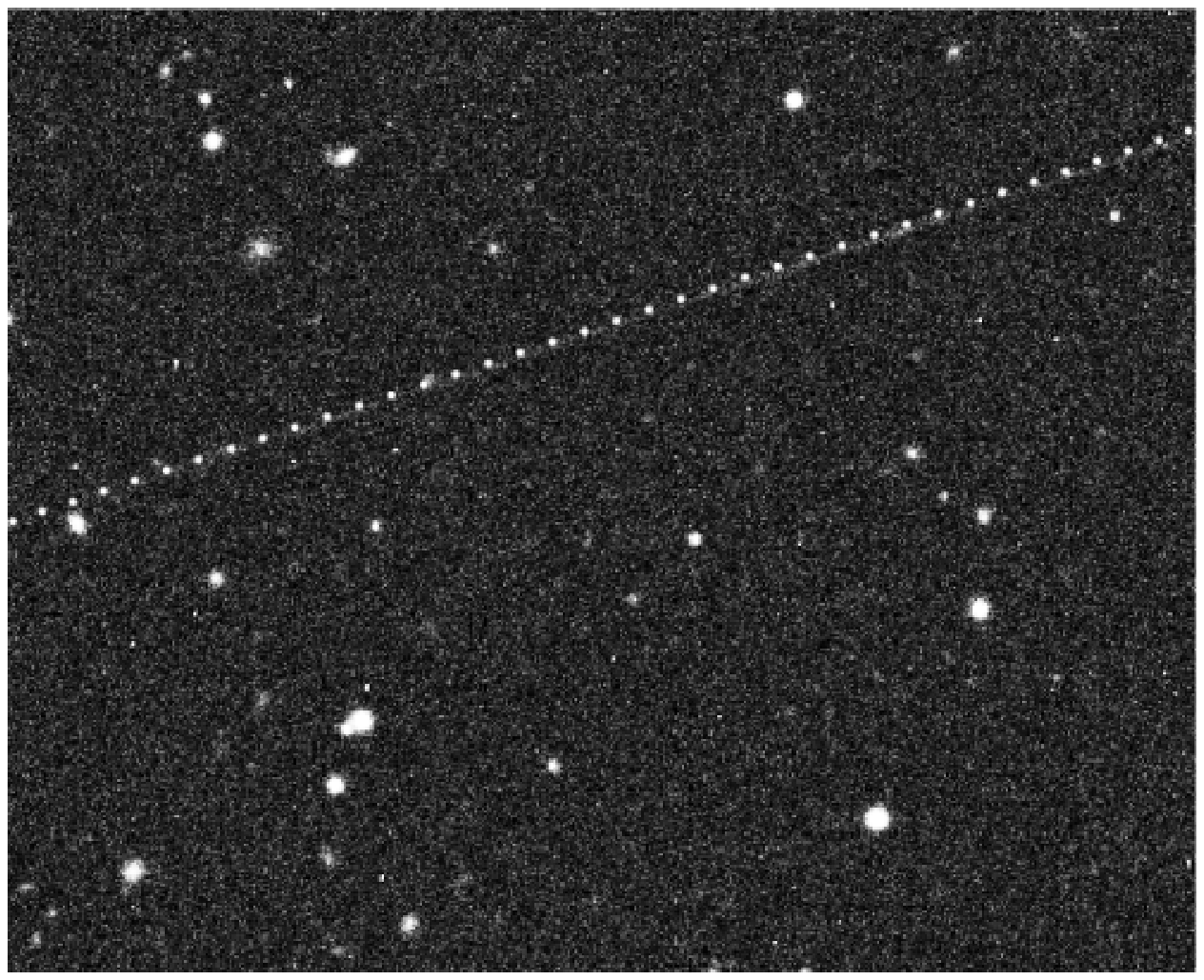}~~~~~~\includegraphics[  height=4.5cm]{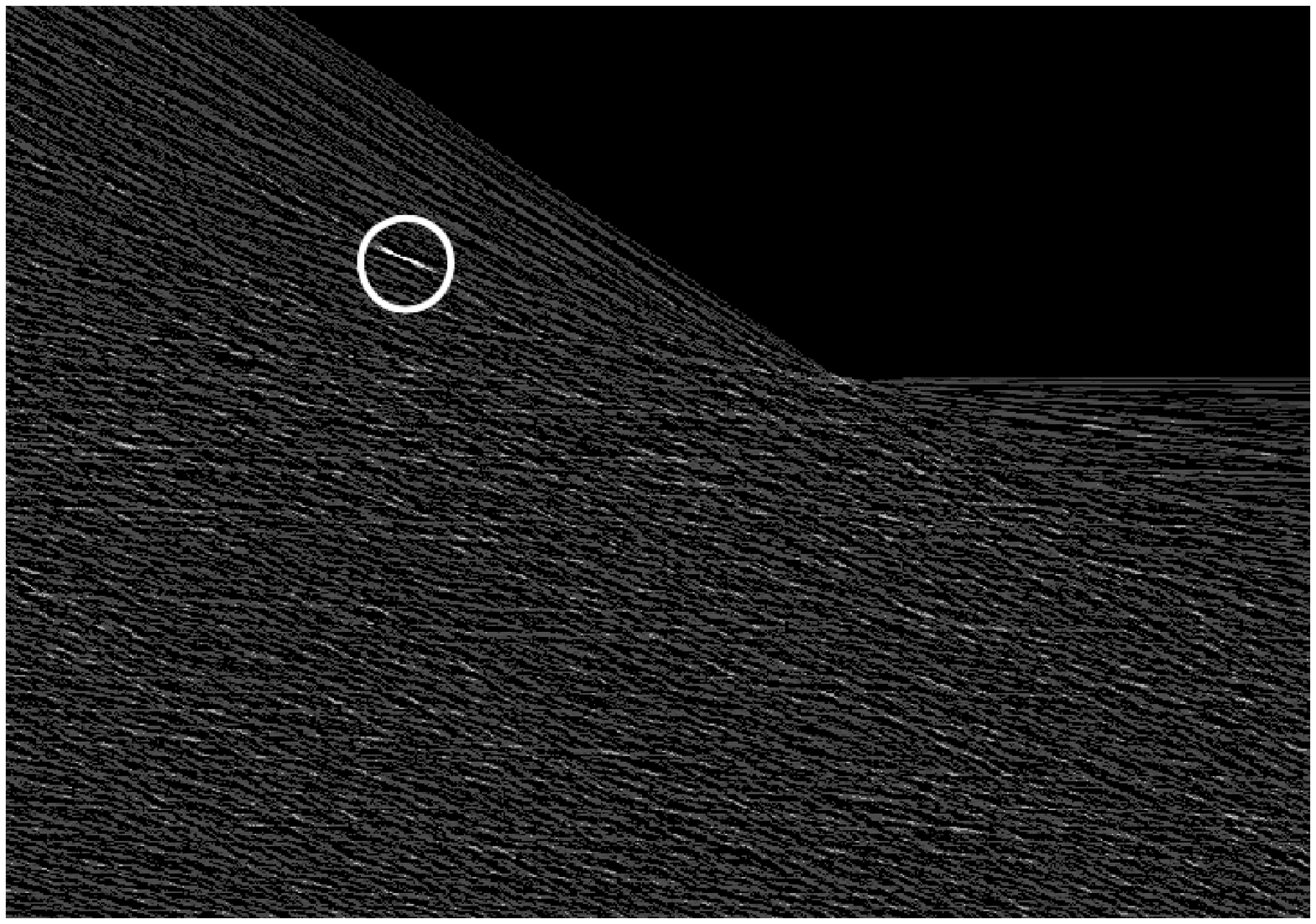}\end{center}

\caption{\label{fig:hough-apli}Detected track in Hough Space, displayed as
a dotted line in the original image.}
\end{figure}

Using the method described above a catalog is extracted from single
exposures considering significant local maxima. This catalog is FHTed,
and a \textit{high-pass} filter is used to remove the constant
response of FHT. Finally, possible tracks are identified on the Hough
filtered image applying a fixed threshold given by
$\frac{\sqrt{N}}{k}$, where $N$ is the number of points in catalog,
and, $k$ is the sensitivity factor.  Tests show that $k=2$ is adequate
for very faint detections without false positives. To illustrate the
method Figure \ref{fig:hough-apli} shows a portion of an image with a
faint satellite track (left panel), where the dashed line represents
the line detected by the procedure on the Hough filtered image (right
panel). This methodology has now been implemented in the EIS pipeline and
it is routinely used with very good results.

\section{General Comments about  the code and Performance}

\begin{figure}[ht]
\begin{center}\includegraphics[  width=130mm]{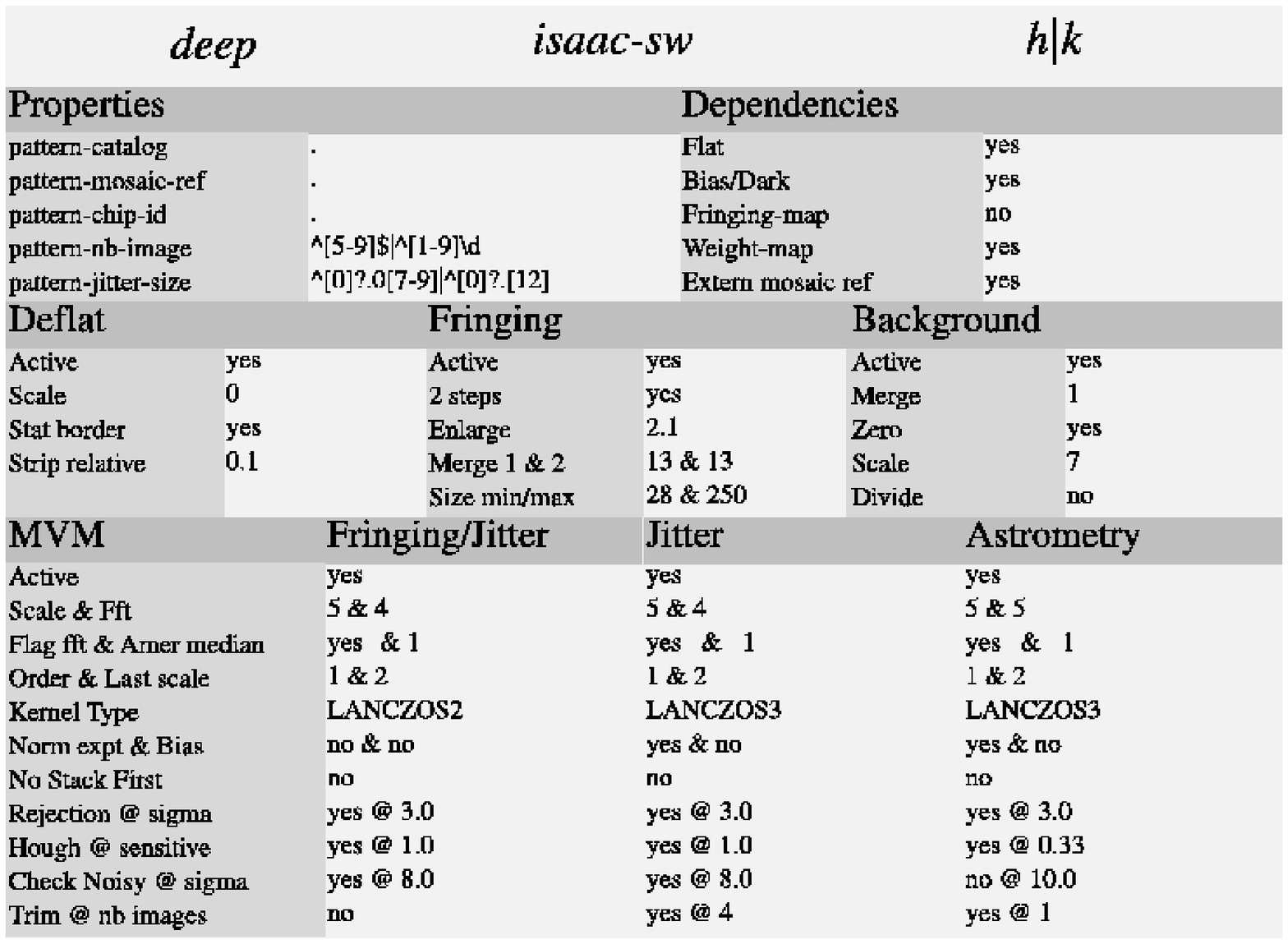}
\label{fig:xslt}
\end{center}

\caption{\label{fig:reduc-mode}An illustration of an HTML page created
from the XML log produced by the system. It displays the main
parameters which define the Reduction Mode used for the IR camera
ISAAC, mounted on one of the ESO VLT telescopes. The reduction mode
shown is used for processing images observed with the filters H or K.
The field properties, for instance, lists the minimum number of
images in the RB and the minimum amplitude of the jitter pattern
required for the execution of this RM. Dependencies lists the master
calibration images needed for reduction.  In this example all steps of
the reduction mode are active, namely: instrumental effect removal;
local background estimation using masks; removal of low-frequency
background; and stacking using relative astrometry, followed by an
absolute registration of the stacked image using a reference grid
specified in the section dependencies. Also shown are possible
settings for the various sub-processes such as: noisy image rejection,
cosmic rays removal and track detection using Hough transform.}
\end{figure}

The software package presented here is an evolution of the so called
multi-resolution visual model or MVM, which consists of basic
functions for operations on images, wavelet transform and image
registration, complemented by open source libraries such as CFITSIO,
FFTW and LIBXML2, among others. The MVM library is written in C++ and
runs on different platforms: SUN-OS, TRUE-64 on an ALPHA computer, and
PC-Linux. Using these basic blocks the package described in this paper
has been developed over the past two years to deal with the specific
needs of astronomical images and consists of 90.000 lines of code.  The
various processes are bound together by employing the Extensible
Mark-up Language (XML) technology which is used to input parameters,
as a way of communicating between different modules, as an internal
database and to store the history of the process. The latter can be
much more comprehensive than it would be possible in the FITS header
of the image.  Furthermore, the associated XML files can be translated
to any format and make their content available in a human-readable
form using suitable style-sheets (XSLT protocol), which can be
processed to produce and display HTML pages. An example is shown in
figure \ref{fig:xslt} where the parameters characterizing a particular reduction
mode are shown.  The figure illustrates the use of a particular
style-sheet which produces a concise description for internal use.  A
more comprehensive style-sheet is used for other users, which gives
more detailed information about each parameter.

By using a common library approach, instead of a concatenation of
stand-alone programs, the package avoids unnecessary access to disk
and the entire reduction is carried out in memory, thereby enhancing
significantly the data rate of the process. This is paramount for
dealing with large data volumes, and is true both for the stand-alone
implementation and the plug-in version for the survey system. In the
latter case, the Python wrapper is only responsible for the
preparation and the administration of the RBs, thereafter launching
the reduction process. The communication between Python and the
library is only done at the beginning and at the end of the process
via the XML configuration and log files produced.

The stand-alone version of the package has been extensively used to
reduce images from a variety of past and ongoing public surveys
allowing thus a fairly robust estimate of its performance. Some examples
are given in Table \ref{table:mvm_resume}, where the survey name, type, instrument,
detector size, number of images and total time required for the
reduction are listed.

\begin{table}
\begin{center}\begin{tabular}{|l|l|l|l|l|l|}
\hline 
Survey name&
 Type of survey&
 Instrument&
 Detector Size&
 \# sci images&
 CPU time\\
\hline
CDF South \cite{cit:arnouts}&
 deep optical&
 WFI @ 2.2&
 8 x 2k x 4k&
 246&
 9 h\\
\hline
CDF South \cite{cit:cdf_ir_vandame}&
 deep IR&
 SOFI @ NTT&
 1 x 1k x 1k&
 4300&
 3h\\
\hline
Pre Flame survey \cite{cit:yazan}&
 stellar optical&
 WFI @ 2.2&
 8 x 2k x 4k&
 26&
 1h\\
\hline
GOOD survey \cite{cit:good}&
 deep IR&
 ISAAC @ VLT&
 1 x 1k x 1k&
 6000&
 4h \\
\hline
\end{tabular}\end{center}

\caption{\label{table:mvm_resume}Examples of the performance of the
system.}
\end{table}

The performance refers to an Alpha DS20E running at 666 MHz with two
CPUs. The values listed above yield a typical reduction data-rate of
$0.5$ Mega-pixels per second for optical images, and $0.4$ Mega-pixels
per second for IR images (which requires the removal of fringes).  The
overhead time for the preparation and the reduction of the calibration
images depends on the instrument, and it is typically $50$ percent of
the reduction time.  The pipeline has also been tested successfully on
other multi-CCD instruments such as: the CFHT 12k camera, and the
SUSI2 instrument mounted on the ESO-NTT telescope.

\section{Conclusion}

In this paper a new image reduction system, incorporating new
technologies and algorithms is presented. It is part of the
effort of the EIS team at ESO to develop an integrated, end-to-end
survey system capable of converting raw images to science-grade
products from multi-wavelength, multi-instrument public surveys. The
system incorporates a number of procedures design to cope with a
variety of particular situations encountered in real observations covering
the optical/infrared domain. The system has a high-throughput ($\sim$
0.5 Mega-pixels/sec) and handles the entire suite of imagers available
at ESO.  It also makes extensive use of XML technology to interface
with the user and to provide the full history of the process. The code
is written in C++ and it is platform independent. The development of
new algorithms is still ongoing (\eg illumination correction) as well
as work aiming at the full integration of the library into the survey
system. Efforts are also made to prepare the required
documentation and user interface to make the stand-alone version of
the image reduction pipeline publicly available.

\begin{acknowledgments}

I would like to express my thanks to A. Bijoui for useful discussions
over the years and to the EIS project and team members for their
support and contribution during the development of this work.

\end{acknowledgments}

\end{document}